\documentclass[preprint,aps,floats,subeqn,showpacs]{revtex4}
\usepackage{epsfig}

\newcommand{\be}{\begin{equation}}
\newcommand{\ee}{\end{equation}}
\newcommand{\bea}{\begin{eqnarray}}
\newcommand{\eea}{\end{eqnarray}}
\newcommand{\bml}{\begin{mathletters}}
\newcommand{\eml}{\end{mathletters}}

\begin{document}

\tighten

\preprint{IUB-TH-047}
\draft




\title{Static solutions of a 6-dimensional Einstein-Yang-Mills model}
\renewcommand{\thefootnote}{\fnsymbol{footnote}}
\author{ Yves Brihaye\footnote{Yves.Brihaye@umh.ac.be}}
\affiliation{Facult\'e des Sciences, Universit\'e de Mons-Hainaut,
7000 Mons, Belgium}

\author{Fabien Clement}
\affiliation{Facult\'e des Sciences, Universit\'e de Mons-Hainaut,
7000 Mons, Belgium}
\author{Betti Hartmann\footnote{b.hartmann@iu-bremen.de}}
\affiliation{School of Engineering and Sciences, International University Bremen (IUB),
28725 Bremen, Germany}

\date{\today}
\setlength{\footnotesep}{0.5\footnotesep}

\begin{abstract}
We study the Einstein-Yang-Mills equations in a 6-dimensional
space-time. We make a self-consistent static, spherically symmetric
ansatz for the gauge fields and the metric. The metric of the
manifold associated with the two extra dimensions contains off-diagonal
terms. The classical equations are solved numerically and several
branches of solutions are constructed. We also present an effective
4-dimensional action from which the equations can equally well be derived.
This action is a standard Einstein-Yang-Mills-Higgs theory extended by three scalar
fields. Two of the scalar fields are interpreted
as dilatons, while the one associated with the off-diagonal term of the
metric induces very specific interactions.

\end{abstract}

\maketitle

\section{Introduction}
In an attempt to unify electrodynamics and general relativity,
Kaluza introduced an extra, a fifth dimension \cite{kaluza} and assumed all
fields to be independent of the extra dimension.
Klein \cite{klein} followed this idea, however, he assumed the fifth dimension
to be compactified on a circle of Planck length. The resulting theory
describes $4$-dimensional Einstein gravity plus Maxwell's equations. 
One of the new fields appearing in this model
is the dilaton, a scalar companion of the metric tensor.
In an analogue way, this field arises in the
low energy effective action of superstring theories and is
associated with the classical scale invariance of these models \cite{maeda}.

Both string theories \cite{pol} as well as so-called ``brane worlds'' 
\cite{brane} assume that space-time possesses more than four dimensions.
In string theories, these extra dimensions are -following the idea of Klein-
compactified on a scale of the Planck length, while in brane worlds,
which assume the Standard model fields to be confined on a 3-brane, they are
large and even infinite. It should thus be interesting to study classical solutions
of non-abelian gauge theories in higher dimensions.

An Einstein-Yang-Mills model in $4+1$ dimensions was studied recently \cite{volkov}.
Assuming the metric and matter fields to be independent on the extra coordinates,
an effective 4-dimensional Einstein-Yang-Mills-Higs-dilaton model appears with
one Higgs triplet and one scalar dilaton. This idea was taken further
to $4+n$ dimensions \cite{bch}, where $n$ Higgs triplets
and $n$ dilatons appear. In contrast to one extra dimension, however,
it appeared that in two or more extra dimensions, constraints
result from the off-diagonal terms of the energy-momentum tensor.
This leads to the fact that only solutions with either
only one non-zero Higgs field or with all Higgs fields constant
are allowed. In this paper, we study an Einstein-Yang-Mills
model in $4+2$ dimensions and we introduce an off-diagonal
term in the metric associated with the extra dimensions in order
to be able to obtain non-trivial solutions. We introduce the model in
Section II and give the Ansatz and equations of motion in Section III.
In Section IV, we give the 4-dimensional effective action
from which the equations of motion can equally well be derived.
In Section V, we present our numerical results and finally, in Section VI 
we give our conclusions.

\section{The Model}

The Einstein-Yang-Mills Lagrangian
in $d=4+2$ dimensions is
given by:

\begin{equation}
\label{action}
  S = \int \Biggl(
    \frac{1}{16 \pi G_{(6)}}R   - 
    \frac{1}{4 e^2}F^a_{M N}F^{a M N}
  \Biggr) \sqrt{g^{(6)}} d^{6} x
\end{equation}
with the SU(2) Yang-Mills field strengths
$F^a_{M N} = \partial_M A^a_N -
 \partial_N A^a_M + \epsilon_{a b c}  A^b_M A^c_N$
, the gauge index
 $a=1,2,3$  and the space-time index
 $M=0,...,5$. $G_{(6)}$ and $e$ denote
respectively the $6$-dimensional Newton's constant and the coupling
constant of the gauge field theory. $G_{(6)}$ is related to the Planck mass
$M_{pl}$ by $G_{(6)}=M_{pl}^{-4}$ and $e^2$ has the dimension of 
$[{\rm length}]^2$.

\section{Ansatz and Equations of motion}
In this paper, we will construct solutions with off-diagonal components
of the metric tensor.  
The Ansatz for the metric then reads:
\begin{eqnarray}
\label{metric}
g_{MN} dx^M
dx^N&=&e^{-(\xi_1+\xi_2~)}
(-A^2 Ndt^2+N^{-1}dr^2 + r^2d \theta^2 + r^2 \sin^2\theta
d\varphi^2)  \nonumber \\
&+& \cosh\left(\frac{J}{2}\right)
\left[e^{2\xi_1}(dx^4)^2+e^{2\xi_2}(dx^5)^2\right]
 + 2 e^{\xi_1+\xi_2} \sinh\left(\frac{J}{2}\right) dx^4 dx^5  \ ,
\end{eqnarray}
where the  functions  $A$, $N$, $J$, $\xi_1$, $\xi_2$ depend
on the variable $r$ only and $N(r)=1-\frac{2m(r)}{r}$.

The determinant of the metric is then given by:
\begin{equation}
\sqrt{-g^{(6)}}=Ar^2\sin{\theta}e^{-(\xi_1+\xi_2)}  \ .
\end{equation}
The $\xi_i$, $i=1,2$ play the role of two scalar dilatons.

The Ansatz for the gauge field reads:
\begin{equation}
A_M^{a}dx^M=A_{\mu}^a dx^{\mu}+ \sum_{k=1}^2 \Phi_k^a dy^k   \  . \
\end{equation}
Note that the $\Phi_j^a$, $j=1,2$ play the role of Higgs fields.  

The spherically symmetric Ansatz is given by \cite{thooft}:
\begin{equation}
{A_r}^a={A_t}^a=0
\ , \end{equation}
\begin{equation}
\label{ansatz1}
{A_{\theta}}^a= (1-K(r)) {e_{\varphi}}^a
\ , \ \ \ \
{A_{\varphi}}^a=- (1-K(r))\sin\theta {e_{\theta}}^a
\ , \end{equation}
\begin{equation}
\label{higgsansatz}
{\Phi}^a_{j}=v H_j(r) {e_r}^a \ \ , \ \ j=1,2 \ ,
\end{equation}
where $v$ is a mass scale
determining the vacuum expectation values of the Higgs fields.

\subsection{Equations of motion}
Using the metric (\ref{metric}), the matter Lagrangian $L_{mat}$ reads:
\begin{eqnarray}\label{reformulation}
L_{mat}=&-&\frac{1}{4}e^{2\xi_1+2\xi_2}F_{\mu\nu}^aF^{\mu\nu,a}
-\frac{1}{2} \cosh\left(\frac{J}{2}\right)
   \left[e^{\xi_2-\xi_1} F_{\mu 4}^a F_4^{\mu,a}
+    e^{\xi_1-\xi_2} F_{\mu 5}^aF_5^{\mu,a}\right]
\nonumber \\
&-& \frac{1}{2}\sinh\left(\frac{J}{2}\right) 
  F_{\mu 5}^a F_4^{\mu,a} 
-\frac{1}{2}e^{-2\xi_1-2\xi_2}F_{45}^aF_{4}^{5,a}  \ \ , \ \ \mu, \nu=0,1,2,3 \ .
\end{eqnarray}
The last term in (\ref{reformulation}) vanishes since
$$F_{45}^aF_{45}^a \propto (\Phi_1^a\times \Phi_2^a)^2=0$$
This is due to the fact that the fields $\Phi_1$, $\Phi_2$ are assumed to
be parallel.

With
\begin{equation}\label{covariant 6D}
F_{\mu (i+3)}^a=\partial_\mu
\Phi_i^a+\varepsilon_{abc}A_\mu^b\Phi_i^c=D_\mu\Phi_i^a \ \ , \ \ i=1,2 \
\end{equation}
the matter Lagrangian $L_{mat}$ reads:
\begin{eqnarray}\label{insertion 6D}
L_{mat}&=&-\frac{1}{4}e^{2\xi_1+2\xi_2}F_{\mu\nu}^aF^{\mu\nu,a} 
\nonumber \\
&-&\frac{1}{2}  \cosh\left(\frac{J}{2}\right)e^{\xi_2-\xi_1}
(D_\mu\Phi_1^aD^\mu\Phi_1^a
+ D_\mu\Phi_2^aD^\mu\Phi_2^a ) \nonumber \\
&-&\frac{1}{2}  \sinh\left(\frac{J}{2}\right)
D_\mu\Phi_1^aD^\mu\Phi_2^a
\end{eqnarray}

Inserting the Ans\"atze (\ref{ansatz1})-(\ref{higgsansatz}) and using

\begin{equation}
x=evr \ \ \ , \ \ \ \mu=evm  \ 
\end{equation}
this reads:
\begin{equation}\label{raccourci}
L_{mat}=-e^{\xi_1+\xi_2}\left[B+\frac{D}{2}+
(\frac{
A_1}{2}+ C_1+\frac{ A_2}{2}+ C_2)\right]
+ A_{12} + 2C_{12}
\end{equation}
with the abbreviations:
\begin{eqnarray}\label{notation}
& A_i&=e^{-2\xi_i}N(H'_i)^2 \cosh\left(\frac{J}{2}\right) \ \ , \ i=1,2  \ , \\
& B&=e^{\xi_1+\xi_2}\frac{N(K')^2}{x^2} \ , \\
& C_i&=e^{-2\xi_i}\frac{K^2H_i^2}{x^2}\cosh\left(\frac{J}{2}\right) \ \ , \ i=1,2, \ , \\
& D&=e^{\xi_1+\xi_2}\frac{(K^2-1)^2}{x^4} \ ,  \\
& A_{12}&=e^{-\xi_1-\xi_2}N H'_1 H'_2\sinh\left(\frac{J}{2}\right) \ , \\
& C_{12}&=e^{-\xi_1-\xi_2}\frac{K^2H_1 H_2}{x^2}\sinh\left(\frac{J}{2}\right) \ .
\end{eqnarray}

\par The non-vanishing components of the energy-momentum tensor
are given by:
\begin{eqnarray}\label{TEI 6D}
&T_0^0&=-\frac{1}{2}e^{\xi_1+\xi_2}\left[2 B+ D+
 A_1+2 C_1+ A_2+2 C_2
-  A_{12} - 2  C_{12}\right] \\
&T_1^1&=e^{\xi_1+\xi_2}\left[ B-\frac{ D}{2}+\frac{
A_1}{2}+\frac{ A_2}{2}- C_1- C_2
 + A_{12} - 2  C_{12}  \right]\\
&T_2^2=T_3^3&=-e^{\xi_1+\xi_2}\left[\frac{ A_1}{2}+\frac{
A_2}{2}-\frac{ D}{2}
 - A_{12} \right]\\
&T_4^4&=-e^{\xi_1+\xi_2}\left[ B+\frac{ D}{2}-\frac{
A_1}{2}+\frac{ A_2}{2}- C_1+ C_2
   \right]\\
&T_5^5&=-e^{\xi_1+\xi_2}\left[ B+\frac{ D}{2}+\frac{
A_1}{2}-\frac{A_2}{2}+ C_1- C_2
   \right]\\
&T_4^5&=-e^{2\xi_1}\left[\tanh\left(\frac{J}{2}\right) ( A_1+2 C_1)
      - \coth\left(\frac{J}{2}\right)( A_{12} + 2  C_{12})
   \right]\\
&T_5^4&=-e^{2\xi_2}\left[\tanh\left(\frac{J}{2}\right)( A_2+2 C_2)
      - \coth\left(\frac{J}{2}\right)( A_{12} + 2  C_{12})
   \right]
\end{eqnarray}


The presence of a off-diagonal term in the metric renders
the components of the Einstein tensor very lengthly and we do not
give them explicitely here. We note that these off-diagonal terms
have to be included for codimension larger than one to avoid 
solutions with trivial Higgs fields. This
is a new feature with respect to \cite{volkov}.

 Five independent Einstein equations can be obtained
 for the five metric functions parametrizing the metric (with $\alpha^2=4\pi G v^2$):

\begin{eqnarray}\label{einstein 1 6D}
\mu'&=&\alpha^2x^2
\left[ B+\frac{ D}{2}+\frac{ A_1}{2}
+C_1+\frac{ A_2}{2}+ C_2
-  A_{12} - 2  C_{12} \right] \nonumber \\
&+&\frac{1}{32}x^2N\left[(J')^2+10 (\xi_1')^2+ 10 (\xi_2')^2+ 12\xi_1'\xi_2'
\right]
\nonumber \\
&+& \frac{1}{16} \cosh\left(\frac{J}{2}\right) 
x^2 N \left(\xi_1'- \xi_2'\right)^2 \ \ ,
\end{eqnarray}

\begin{eqnarray}\label{einstein 2 6D}
A'&=&\frac{\alpha^2Ax}{N}
\left[2 B+ A_1+ A_2-2  A_{12}\right]
+\frac{1}{16}Ax\left[(J')^2+ 10(\xi_1')^2 \right. \nonumber \\
&+& \left. 10(\xi_2')^2+12\xi_1'\xi_2'
+ 2 \cosh\left(\frac{J}{2}\right) x^2 N (\xi_1'- \xi_2')^2\right] \ \ ,
\end{eqnarray}\par

\begin{eqnarray}\label{einstein 3 6D}
&\xi_1''&=
- \tanh\left(\frac{J}{2}\right) J' (\xi_1' - \xi_2') 
- \xi_1' \frac{1+N}{N}  \nonumber \\
&+& \alpha^2 \left[
\hat T_{22} \frac{2 \xi_1'}{xN}
+\hat T_{44} e^{-\xi_2-3\xi_1}
\frac{\cosh(\frac{J}{2})(x\xi_1'-1)+ x\xi_1'-3}{2N\cosh(\frac{J}{2})}
 \nonumber \right. \\
&+&\left. \hat T_{55} e^{-\xi_1-3\xi_2}
\frac{\cosh\left(\frac{J}{2}\right)(x\xi_1'-1)+ x\xi_1'+1}{2N\cosh(\frac{J}{2})}
+\hat T_{54} e^{-2\xi_1-2\xi_2}
\frac{2\sinh(\frac{J}{2})(1 - x\xi_1')}{N}
\right] \ \ ,
\end{eqnarray}


\begin{eqnarray}\label{einstein 5 6D}
 J'' &=& (\xi_1'-\xi_2')^2 - J' \frac{1+N}{x N}
  \nonumber \\
 &+& \alpha^2 \left[
       \hat T_{2 2} \frac{2 J'}{xN}
      +\hat T_{4 4} e^{-3 \xi_1 - \xi_2}
      \frac{\cosh(\frac{J}{2})x J' + 4 \sinh(\frac{J}{2})}{N} \right.
      \nonumber \\
       &+&\left. \hat T_{5 5} e^{-3 \xi_2 - \xi_1}
       \frac{\cosh(\frac{J}{2})x J' + 4 \sinh(\frac{J}{2})}{N}
       -2\hat T_{5 4} e^{-2 \xi_2 - 2\xi_1}
       \frac{4 \cosh(\frac{J}{2}) +  x J'\sinh(\frac{J}{2})}{N}
      \right] \ ,
\end{eqnarray}
where we use the abbreviation
\begin{equation}
\hat T_{M N} \equiv T_{M N}- \frac{1}{4}g_{M N} 
T_K^K \ \ , \ \ M,N,K=0,1,2,3,4,5 \ ,
\end{equation}
and the prime denotes the derivative with respect to $x$.
Note that the equation for $\xi_2$ can be obtained for (\ref{einstein 3 6D})
by exchanging $\xi_1\leftrightarrow \xi_2$.

Finally the variation of the action (\ref{action})
with respect to the matter fields leads to three 
differential equations
for the functions
$K$, $H_1$ and $H_2$:
\begin{eqnarray}
\label{mat 1 6D}
(K'e^{\xi_1+\xi_2}AN)'&=&AK\left[e^{\xi_1+\xi_2}\frac{(K^2-1)}{x^2}+
\cosh\left(\frac{J}{2}\right)(e^{-2\xi_1}(H_1)^2+e^{-2\xi_2}(H_2)^2)\right]  \nonumber \\
&-& 2 A H_1 H_2 K e^{-\xi_1-\xi_2} \sinh\left(\frac{J}{2}\right) \ ,
\end{eqnarray}
\begin{eqnarray}\label{mat 2 6D}
\frac{e^{2 \xi_a}}{x^2 A N} (e^{-2\xi_a}ANx^2H_a')'&=&
2 \frac{1}{x^2 N} K^2 H_a
+ \frac{1}{2} e^{\xi_a-\xi_b} H_b'\left(J'-(\xi_a-\xi_b) 
\sinh\left(\frac{J}{2}\right)\right) \nonumber \\
&-& \frac{1}{2}\left(1- \cosh
\left(\frac{J}{2}\right)\right) (\xi_a'-\xi_b')H_a'  \ ,
\end{eqnarray}
where $a,b = 1,2$ and $a \neq b$.

\subsection{Boundary conditions}

 We will study globally regular, asymptotically flat 
solutions of the system above. This implies the following boundary conditions:
\begin{equation}
K(0)=1 \ , \ \ H_j(0)=0  \ , \ 
 \mu(0)=0 \ , \ \partial_x J|_{x=0}=0 \ , \  \partial_{x}\xi_j|_{x=0}=0 
\ , \ j=1,2   \label{bc1}
\end{equation}
at the origin and 
 \begin{equation}
K(\infty)=0 \ ,  \ H_j(\infty)=c_j \ ,  \ A(\infty)=1 \ , \ J(\infty)=0 \ , \
\xi_j(\infty)=0 \ , \ j=1,2  
\label{bc2} \end{equation}
at infinity.

\section{The 4-dimensional effective action}
The above equations can be obtained equally from the following 4-dimensional
effective action:

 \begin{eqnarray}
 \label{lageffj}
&L_{mat}&=-\frac{1}{4}e^{2\kappa(\Psi_1+\Psi_2)}F_{\mu\nu}^aF^{\mu\nu,a}
           -\frac{1}{2}e^{-4\kappa\Psi_1}\cosh{\left(\frac{J}{2}\right)}
           (D_\mu\Phi_1^a)(D^\mu\Phi_1^a)\nonumber\\
&&-\frac{1}{2}e^{-4\kappa\Psi_2}\cosh{\left(\frac{J}{2}\right)}
(D_\mu\Phi_2^a)(D^\mu\Phi_2^a)
+e^{-2\kappa(\Psi_1+\Psi_2)}\sinh{\left(\frac{J}{2}\right)}
(D_\mu\Phi_1^a)(D^\mu\Phi_2^a)\nonumber\\
&  &-\frac{5}{12}[~(\partial_\mu\Psi_1)(\partial^\mu\Psi_1)
        +(\partial_\mu\Psi_2)(\partial^\mu\Psi_2)~]\nonumber\\
&   &-\frac{1}{2}(\partial_\mu\Psi_1)(\partial^\mu\Psi_2)
-\frac{1}{12}\cosh{(J)}~[~(\partial_\mu\Psi_1)(\partial^\mu\Psi_1)
+(\partial_\mu\Psi_2)(\partial^\mu\Psi_2)\nonumber\\
&&-2(\partial_\mu\Psi_1)(\partial^\mu\Psi_2)~]-\frac{1}{32}(\partial_\mu
J)(\partial^\mu J)
 \end{eqnarray}
provided the following identification is done
\begin{equation}
\xi_i=2\kappa \Psi_i \ \ , \ \ i=1,2 \ \ , \ \ \ \ \   \alpha^2=3\kappa^2 \ .
\end{equation}
Note that $\kappa$ is a new coupling constant (the ``dilaton'' coupling)
appearing in the effective action. 
The function $J$ appears as a new dynamical scalar field in the effective
action.

\section{Discussion of numerical results}

\begin{figure}
\centering
\epsfysize=20cm
\mbox{\epsffile{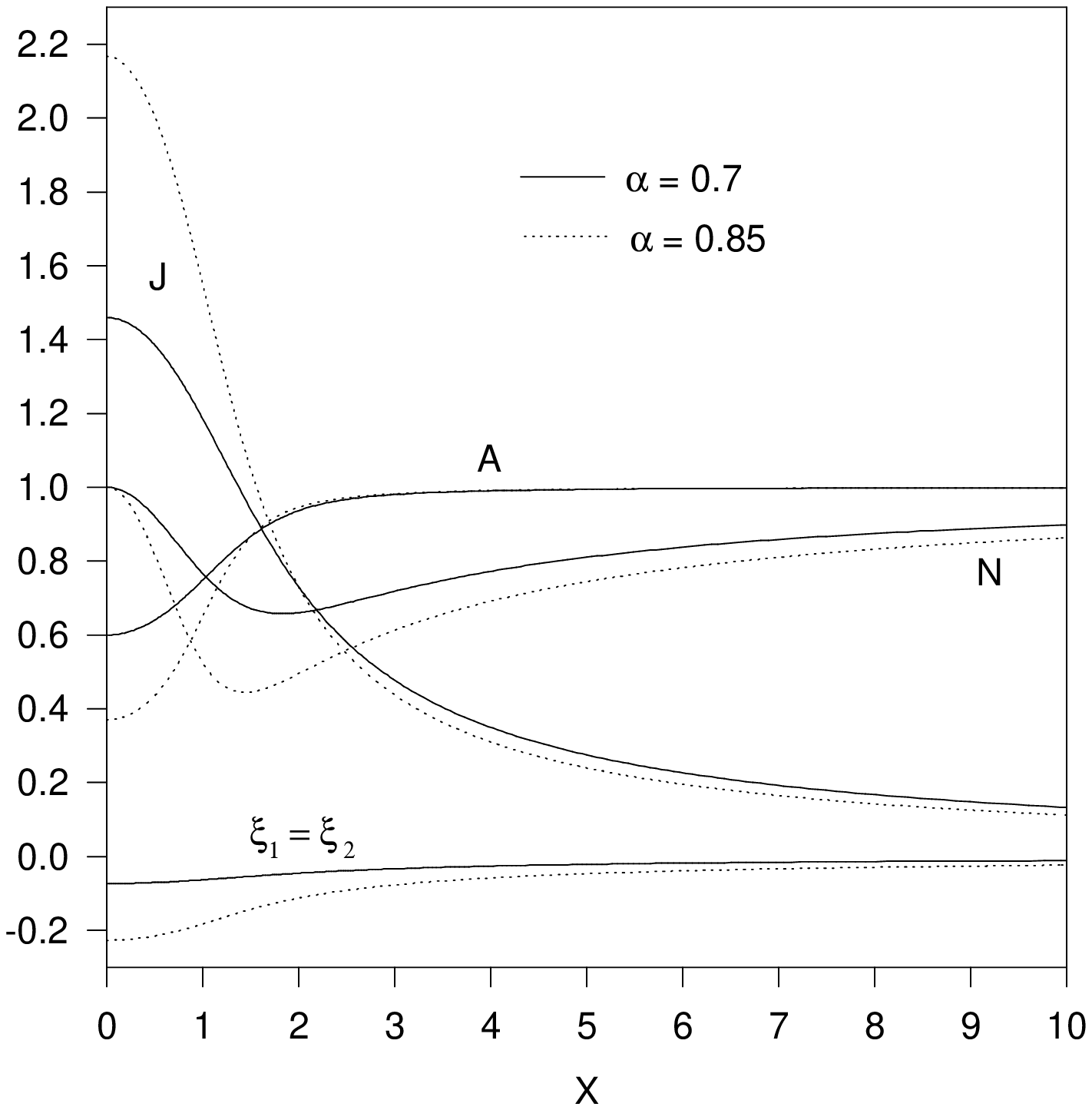}}
\caption{\label{fi45f1} The profiles of the metric functions
$A(x)$, $J(x)$, $N(x)$, $\xi_1(x)=\xi_2(x)\equiv\xi(x)$ are shown for two different values of
$\alpha$.}
\end{figure}

The system of non-linear differential equations can only be solved numerically.
We have solved this system for numerous values of $\alpha$. For the numerical analysis
we assume $H_1(\infty)=H_2(\infty)=1$, but we believe that the pattern
of solutions remains qualitatively equal for $H_1(\infty)
>  H_2(\infty) >0$.
In the limit $\alpha=0$ the metric is flat (Minkowski metric) 
(implying $J(x)=\xi_1(x)=\xi_2(x)=0$) and the solution
of the equations is the 't Hooft-Polyakov magnetic monopole \cite{thooft}.
Due to the presence of two -linearly superposed- Higgs fields,
the flat solution has mass $\sqrt{2}$, which is confirmed by our numerical
results.
For $\alpha > 0$ the solution is progressively deformed by gravity,
the fields $J(x)$, $\xi_1(x) = \xi_2(x)\equiv \xi(x)$, $A(x)$, $N(x)$ 
become non-trivial.
In particular, $N(x)$ develops a local minimum at some finite radius
$N(x_m)=N_m$,
$0<A(0)<1$ represents the minimum of the function $A(x)$, the dilaton
field $\xi(x)$ is negative and monotonically increasing, while the function $J(x)$ is
positive and decreases monotonically from $J(0)>0$ to $J(\infty)=0$. 
The profiles of the different functions is presented in Fig.\ref{fi45f1}
for two different values of $\alpha$. This figure shows that
$\vert J(x)\vert$ is larger than $\vert\xi_1\vert=\vert\xi_2\vert\equiv\vert\xi(x)\vert$.
 
\begin{figure}
\centering
\epsfysize=20cm
\mbox{\epsffile{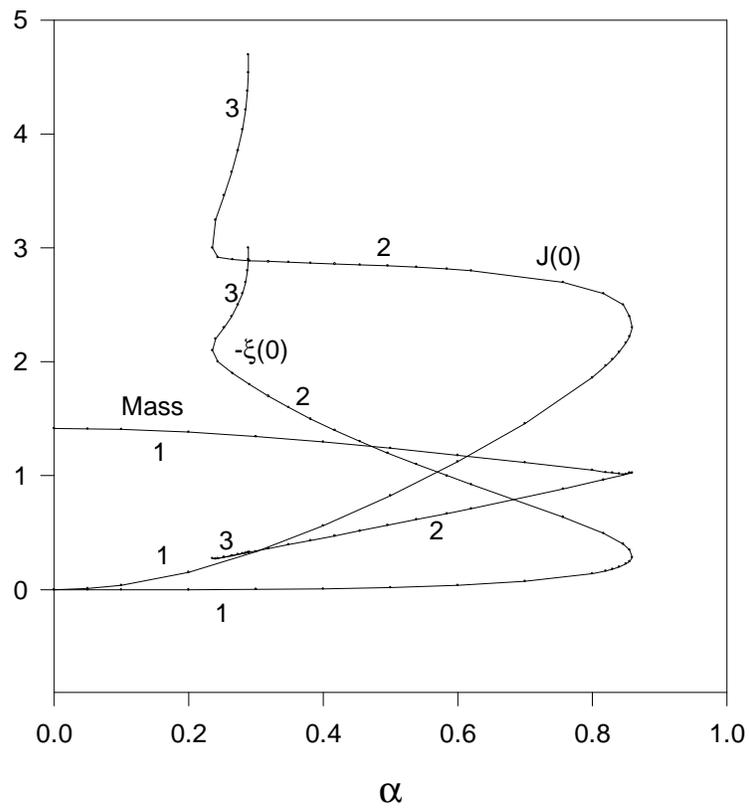}}
\caption{\label{fi45f2} The dependence of the values 
$J(0)$ and $-\xi_1(0)=-\xi_2(0)\equiv-\xi(x)$ as well as
the ADM mass are shown as functions of $\alpha$.}
\end{figure}

\begin{figure}
\centering
\epsfysize=20cm
\mbox{\epsffile{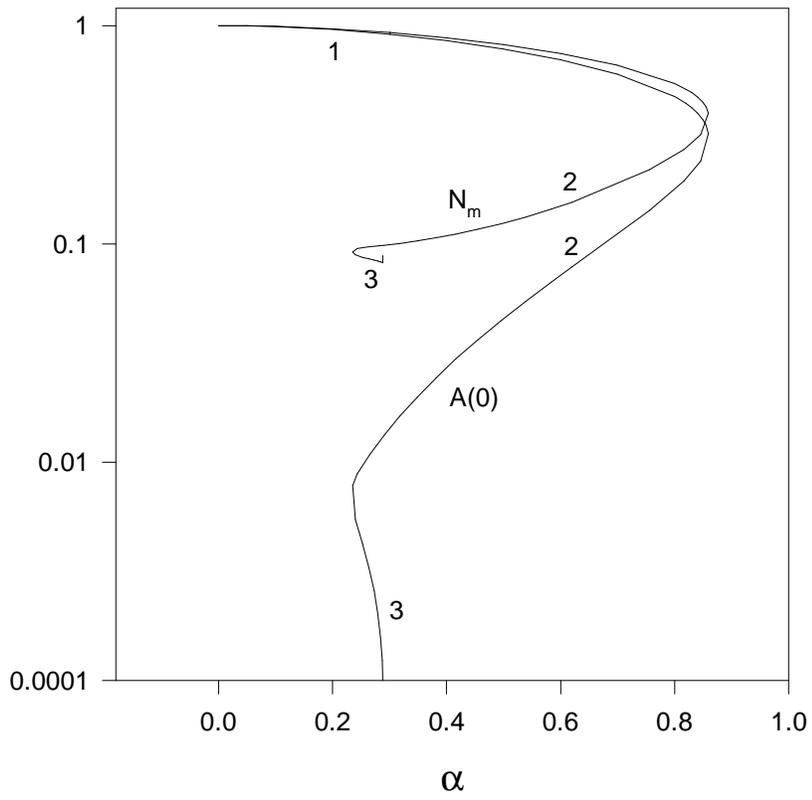}}
\caption{\label{fi45f4}The dependence of $N(x_{min})=N_m$ and $A(0)$
are shown as functions of $\alpha$.}
\end{figure}

The values $\xi(0),J(0)$ and $A(0),N_m$ are plotted as functions of
$\alpha$ 
in Fig. \ref{fi45f2} and Fig. \ref{fi45f4}, respectively. 
The branch which arises from the gravitational deformation
of the flat monopole
is indexed by the label ``1'' in the figures. In Fig.\ref{fi45f2}
we also plot the ADM mass of the solution, which decreases with
increasing $\alpha$.

The figures further demonstrate that the branch ``1'' of solutions does not
exist for arbitrary large values of $\alpha$. Rather, the solutions
exist only for $0\leq \alpha \leq \alpha_{max}$ with 
$\alpha_{max} \approx 0.859$. In addition, another branch of solutions
exists for $ \alpha \leq  \alpha_{max}$. More precisely
, we have managed to construct a second branch of solution (labelled ``2'' in 
the figures) for $\alpha_{1} \leq \alpha \leq \alpha_{max}$ (with
$\alpha_1 \approx 0.235$). Again, starting from $\alpha_1$ and increasing
$\alpha$, our 
numerical analysis strongly suggests the existence of a third branch
labelled ``3''. Note that the ADM mass of the solutions of 
branch ``3'' is very close to that of the solutions of branch ``2''.
This is why the two branches cannot be distinguished on the plot 
as far as their ADM mass is concerned.  
Continuing this construction, there probably exist
a large (possibly infinite) number of branches on smaller and smaller
intervals concentrated around the value $\alpha \approx 0.288$.
This pattern is very reminiscent to the one occuring in the 
case with $d=4+1$ first discussed by Volkov \cite{volkov}. 
However the behaviour of the mass
of the solutions in the $d=4+1$ case and in the $d=4+2$ cases is considerably
different. Indeed, for
$d=4+1$, the mass of the solutions on the branches ``2'',``3'' 
is higher than the mass of the solution on the branch ``1''; 
in the present case, it is the
contrary. Said in other words, for 
$\alpha_1 \leq \alpha \leq  \alpha_{max}$
the solution with the lowest mass is the one of branch ``2''.

At first glance this result may appear paradoxal, since it is known
that the monopole is topologically stable. However, we believe that
this peculiar behaviour of the mass is deeply connected to the 
direct coupling between the two Higgs fields in the effective action.
Roughly speaking the two Higgs fields are coupled with a factor
$-\sinh(J)$ (remember that this term is absent in absence of gravity
$J(x)=0$).
For $J > 0$ it turns out that the three contributions due to the
terms quadratic in the derivative of the Higgs fields have
a tendency to compensate.
For the solutions of the second branch, the dilatons $\xi_1(x)=\xi_2(x)$ and
the  function $J(x)$ deviate stronger from
their flat space-time values than on the branch ``1''. Specifically,
$J(x)$ becomes large enough  to 
dimnish considerably the contribution of the Higgs field to the mass.
As a consequence, the mass becomes smaller than the mass on the first
branch. 

Analyzing this result from the point of view of catastrophe theory
it it tempting to conclude that the solutions on the branch ``1'' are
unstable (they are sphaleron-like)
and it is very likely that the monopole looses its 
non-trivial topology and its status
of being a local minimum when it is coupled to the other fields appearing  
in the 4-dimensional effective model. There are
$k$ negative -unstable- modes while the solution of branch ``2''
have $k-1$ unstable modes.  A direct analysis of the stability should
definitely confirm this conclusion and provide the value of $k$.

\section{Conclusions}
The dimensional reduction of the Einstein-Yang-Mills
theory from $4+n$ dimensions down to the standard $4$ dimensions 
is an interesting
problem. The main source of uncertainty is the way the
fields depend on the extra dimensions.
One can adopt, for the simplest case, the point of view that,
in an appropriate gauge and with appropriate variables,
the gauge and metric fields are independent on the extra
coordinates.
A 4-dimensional effective action which encodes all
the effects of the extra dimension into a more or less
``conventional'' field theory can then be constructed.

In the case of $n=1$, it was shown that
the effective action is a Georgi-Glashow model
appropriately coupled to one dilaton field \cite{volkov}. 
Here we have studied
in details the case $n=2$. 
One of the main differences in comparison to the
$n=1$ case is that the Einstein-Yang-Mills equations are consistent
only if the metric has off-diagonal terms in the subspace
of the extra dimensions. The corresponding effective action
is a standard Einstein-Yang-Mills action 
supplemented by two Higgs triplets
and three  scalar fields, originating from the parametrization
of the metric in the codimensions. The function corresponding to
the off-diagonal term of the metric plays a very important role
in the effective action, largely determining an interaction
between the two Higgs fields.
All these fields acquire kinetic parts as well as specific interactions.

The equations of motion can then be studied numerically.
We  have constructed several branches of
classical solutions which can be seen as
deformations of a magnetic monopole.

Several possibilities of extension of the result can be investigated,
namely (1) the construction of non-abelian
black hole, (2) the stability analysis
of the different branches of regular solutions, (3) the addition
of higher order terms in the Yang-Mills action (``Born-Infeld'' terms)
and (4) the construction of axially symmetric solutions.

  \newpage

 \newpage
\newpage

\end{document}